\documentclass[preprint,showpacs,preprintnumbers,amsmath,amssymb]{revtex4}
\usepackage{booktabs}
\usepackage{mathrsfs}
\usepackage{epsfig}
\usepackage{graphicx}
\usepackage{dcolumn}
\usepackage{bm}
\usepackage{amsmath}
\let\jnfont=\rm
\def\NPB#1,{{\jnfont Nucl.\ Phys.\ B }{\bf #1},}
\def\PLB#1,{{\jnfont Phys.\ Lett.\ B }{\bf #1},}
\def\EPJC#1,{{\jnfont Euro.\ Phys.\ J.\ C }{\bf #1},}
\def\PRD#1,{{\jnfont Phys.\ Rev.\ D }{\bf #1},}
\def\PRL#1,{{\jnfont Phys.\ Rev.\ Lett.\ }{\bf #1},}
\def\MPLA#1,{{\jnfont Mod.\ Phys.\ Lett.\ A }{\bf #1},}
\def\JPG#1,{{\jnfont J.\ Phys.\ G}{\bf #1},}
\def\CTP#1,{{\jnfont Commun.\ Theor.\ Phys.\ }{\bf #1},}
\def\JHEP#1,{{\jnfont J. High \ Ener.\  Phys.}{bf #1},}
\def\RMP#1,{{\jnfont  Rev. Mod. Phys.}{bf #1},}

\def\Rv{\not{\hbox{\kern-1pt $R$}}}
\def\p{\not{\hbox{\kern-3pt $p$}}}

\begin{document}
\preprint{\parbox{1.2in}{\noindent arXiv:1002.0659}}

\title{Probing topcolor-assisted technicolor from
       lepton flavor violating processes in photon-photon collision at ILC}

\author{Guo-Li Liu}\thanks{E-mail:guoliliu@zzu.edu.cn}
\affiliation{Physics Department, Zhengzhou University, Henan,
450001, China; \\
Kavli Institute for Theoretical Physics China, Academia Sinica,
Beijing 100190, China}


\begin{abstract}
In topcolor-assisted technicolor models (TC2) the hitherto
unconstrained lepton flavor mixing induced by the new gauge boson
$Z'$ will lead to the lepton flavor violating productions of
$\tau\bar \mu$, $\tau\bar e$ and $\mu\bar e$ in photon-photon
collision at the proposed International Linear Collider (ILC).
Through a comparative analysis of these processes, we find that the
better channels to probe the TC2 is the production of $\tau\bar \mu$
or $\tau\bar e$ which occurs at a much higher rate than $\mu\bar e$
production due to the large mixing angle and the large flavor
changing coupling, and may reach the detectable level of the ILC for
a large part of the parameter space. Since the rates predicted by
the Standard Model are far below the detectable level, these
processes may serve as a sensitive probe for the TC2 model.
\end{abstract}

 \maketitle
\section{\bf Introduction}

Since in the Standard Model (SM) the lepton flavor violating (LFV)
interactions are extremely suppressed, any observation of the LFV
processes would serve as a robust evidence for new physics beyond
the SM. These LFV processes, which have been searched in various
experiments \cite{exp2,exp3,exp4-l-lr}, can be greatly enhanced in
new physics models like supersymmetry \cite{LFC-1,rrmutau-susy} and
the topcolor-assisted models (TC2)
\cite{tc2-rev,lfv-tc2,eemutau-tc2}. Such enhancement can be several
orders to make them potentially accessible at future collider
experiments.

Due to its rather clean environment, the proposed International Linear
Collider (ILC) will be an ideal machine to probe new physics. In
such a collider, in addition to $e^+ e^-$ collision, we can also
realize $\gamma \gamma$ collision with the photon beams generated by the backward
Compton scattering of incident electron- and laser-beams. The LFV interactions
in TC2 model will induce various processes at the ILC, such as
the productions of $\tau\bar \mu$, $\tau \bar e$ and $\mu \bar e$
via $e^+ e^-$. 
 It is noticeable that the productions of  $\tau\bar \mu$, $\tau \bar e$ and
$\mu \bar e$ in $\gamma \gamma$ collision have not been studied in
the framework of the TC2. Such LFV productions in $\gamma \gamma$
collision may be more important than in $e^+ e^-$ collision
\cite{eemutau-tc2} collision since these productions are a good
probe for new physics because it is essentially free of any SM
irreducible background. 
 It is also noticeable that all these LFV
processes at the ILC involve the same part of the parameter space of
the TC2.
 Therefore, it is necessary to comparatively study all these processes
to find out which process is best to probe the TC2 model.

We in this work will study the LFV processes $\gamma\gamma \to \ell_i\bar \ell_j$
($i\neq j$ and $\ell_i = e,~\mu~\tau $) induced by the extra $U(1)$ gauge boson
$Z'$ in TC2 models. We calculate the production rates
to figure out if they can reach the sensitivity of the photon-photon collision
of the ILC.

The work is organized as follows. We will briefly discuss the TC2
model in Section II, giving the new couplings which will be involved
in our calculation. In Section III and IV we give the calculation
results and compare with the results in the SUSY. Finally, the
conclusion is given in Section V.

\section{About TC2 model}
There are many kinds of new physics scenarios predicting new particles, which
can lead to significant LFV signals. For example, in the minimal
supersymmetric SM, a large $\nu_\mu - \nu_\tau$ mixing leads to clear LFV
signals in slepton and sneutrino production and in the decays of neutralinos
and charginos into sleptons and sneutrinos at hadron colliders and lepton
colliders \cite{vv-mixing}. The non-universal U(1) gauge bosons $Z'$, which are
prediced by
various specific models beyond the SM, can lead to the large tree-level flavor
changing(FC) couplings. Thus, these new particles may have significant
contributions to some LFV processes \cite{z'couple}.

The key feature of TC2 models \cite{tc2-rev} and flavor-universal
TC2 models \cite{K_1} is that the large top quark mass is
mainly generated by topcolor interactions at a scale of order
1 TeV . The topcolor interactions may be flavor non-universal (as in
TC2 models) or flavor-universal (as in flavor-universal TC2 models).
However, to tilt the chiral condensation in the $t\bar{t}$ direction
and not form a $b\bar{b}$ condensation, all of these models need a
non-universal extended hypercharge group $U(1)$. Thus, the existence
of the extra $U(1)$ gauge bosons $Z^{\prime}$ is predicted. These
new particles treat the third generation fermions (quarks and
leptons) differently from those in the first and second generations,
namely, couple preferentially to the third generation fermions.
After the mass diagonalization from the flavor eigenbasis into the
mass eigenbasis, these new particles lead to tree-level FC
couplings. The flavor-diagonal couplings of the extra $U(1)$ gauge
bosons $Z^{\prime}$ to ordinary fermions, which are related to our
calculation, can be written as \cite{tc2-rev,exp-tc2}:
\begin{equation}
{\cal L}=-\frac{1}{2}g_{1}\{\tan\theta^{\prime}\{(\bar{e}_{L}\gamma^{\mu}
         e_{L}+2\bar{e}_{R}\gamma^{\mu}e_{R} + \bar{\mu}_{L}\gamma^{\mu}
         \mu_{L}+2\bar{\mu}_{R}\gamma^{\mu}\mu_{R}) +
        \cot\theta' (\bar{\tau}_{L}\gamma^{\mu}
         \tau_{L}+2\bar{\tau}_{R}\gamma^{\mu}\tau_{R})\}\cdot Z'_{\mu}
\end{equation}
where $g_{1}$ is the ordinary hypercharge gauge coupling constant.
$\theta^{\prime}$ is the mixing angle and
$\tan\theta^{\prime}=\frac{g_{1}}{2\sqrt{\pi K_{1}}}$ where $K_1$ is
the coupling constant.

The flavor-changing couplings of the extra $U(1)$ gauge bosons
$Z^{\prime}$ to ordinary fermions, which are related to our
calculation, are given in the followings: \cite{tc2-rev,exp-tc2}:
\begin{equation}
{\cal L}=-\frac{1}{2}g_{1}\{K_{\mu e}(\bar{e}_{L}\gamma^{\mu}
         \mu_{L}+2\bar{e}_{R}\gamma^{\mu}\mu_{R})+k_{\tau\mu}(\bar{\tau}_{L}
         \gamma^{\mu}\mu_{L}+2\bar{\tau}_{R}\gamma^{\mu}\mu_{R})
+k_{\tau e}(\bar{\tau}_{L}
         \gamma^{\mu}e_{L}+2\bar{\tau}_{R}\gamma^{\mu}e_{R})  \}\cdot Z'_{\mu},
\end{equation}

 where $k_{\mu e}$, $k_{\tau e}$ and $k_{\tau\mu}$ are the flavor mixing
 factors. Since the new gauge boson $Z'$ couples preferentially to the third
 generation, the factor $K_{\mu e}$ are negligibly small, so in the
 following estimation, we will neglect the $\mu- e$ mixing, and consider only
 the flavor changing coupling processes $\gamma\gamma \to \tau\bar\mu$ and $\gamma\gamma \to
 \tau \bar e$.

Note that the difference between the $Z'\tau\bar\mu$ and $Z'\tau\bar
 e$ couplings lies only in the flavor mixing factor $K_{\tau\mu}$
and $K_{\tau e}$ and the masses of the final state leptons. Since the
non-universal gauge boson $Z'$ treats the fermions in the third generation
differently from those in the first and second generations and treats the
 fermions in the first same as those in the second generation, so in the
following calculation, we will assume $K_{\tau\mu} = K_{\tau e}$.
Then what makes the discrepancy of the cross sections of the two
channels $\gamma\gamma \to \tau\bar \mu$ and $\gamma\gamma \to \tau
\bar e$ is only the masses of the final state particles. Considering
the large mass $M_{Z'}> 1$ TeV, for simplicity, we will take
$M_\tau=M_\mu=M_e=0$ in the following discussion, i.e., assuming the
cross sections of the two channels $\gamma\gamma \to \tau\bar \mu$
and $\gamma\gamma \to \tau\bar e$ are equal to
 each other.

\section{Calculation}

\begin{figure}[tbh]
\begin{center}
\epsfig{file=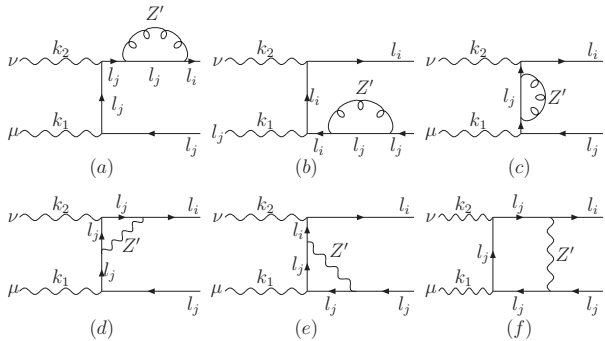,width=8cm} \caption{ Feynman diagrams
contributing to the process $\gamma\gamma \to \ell_i\ell_j$ in TC2
models.} \label{fig1}
\end{center}
\end{figure}

 The Feynman diagram of the LFV processes $ \gamma\gamma \to
\ell_i\ell_j$ ($i\neq j$ and $\ell_i = e,~\mu,~\tau$) induced by the
extra U(1) $Z'$ is shown in
 Fig.~\ref{fig1}. There are only t- and u- channel contributions,
the latter not shown in Fig~\ref{fig1}, but We can calculate them by
exchanging the two photons.

Note that there is no s-channel contribution to the LFV processes.
As we know, in the SM production of on-shell $Z$ boson at a
photon-photon collider (or $Z$ decays into $\gamma\gamma$) is
strictly forbidden by angular momentum conservation and Bose
statistics, which is the predict of the famous Laudau-Yang Theorem.
This theorem is still effective to our case, since that the two real
photons cannot be in a state with angular momentum $J=1$ regardless
of on-shell or off-shell bosons, so the s-channel contribution with
two real photons to an extra $Z'$ vanishes automatically.


   The electroweak gauge bosons $\gamma$ and $Z$ can not couple to
 $\tau\bar e$, $\mu\bar e$ and $\tau\bar \mu$, so we need not consider
the interference effects  between the $\gamma$, $Z$ and $Z^{\prime}$
on the cross section of the process $\gamma\gamma \to \ell_i
\bar\ell_j$($i\neq j$ and $\ell_i = e,~\mu,~\tau$). In TC2 models
the gauge invariant amplitude of $\gamma\gamma \to \tau\bar\mu(\bar
e)$ induced by the extra boson $Z'$ is given by
\begin{eqnarray}
{\cal M}=\frac{1}{2} ~\bar{u}_\tau\Gamma^{\mu\nu}P_Lv_\mu
~\epsilon_\mu(\lambda_1)\epsilon_\nu(\lambda_2)
\end{eqnarray}
where the $\Gamma^{\mu\nu}$ is defined same as that in
\cite{rrtc}. These amplitudes contain the Passarino-Veltman
one-loop functions, which are calculated by using LoopTools
\cite{Hahn}.

Since the photon beams in $\gamma\gamma$ collision are generated
by the backward Compton scattering of the incident electron- and
the laser-beam, the events number is obtained by convoluting the
cross section of $\gamma\gamma$ collision with the photon beam
luminosity distribution:
\begin{eqnarray}
N_{\gamma \gamma \to \ell_i \bar\ell_j}&=&\int d\sqrt{s_{\gamma\gamma}}
  \frac{d\cal L_{\gamma\gamma}}{d\sqrt{s_{\gamma\gamma}}}
  \hat{\sigma}_{\gamma \gamma \to \ell_i \bar\ell_j}(s_{\gamma\gamma})
  \equiv{\cal L}_{e^{+}e^{-}}\sigma_{\gamma \gamma \to \ell_{i} \bar\ell_{j}}(s)
\end{eqnarray}
where $d{\cal L}_{\gamma\gamma}$/$d\sqrt{s}_{\gamma\gamma}$ is the photon-beam luminosity
distribution and $\sigma_{\gamma \gamma \to \ell_i \bar\ell_j}(s)$ (
$s$ is the squared center-of-mass energy of $e^{+}e^{-}$ collision) is defined as
the effective cross section of $\gamma \gamma \to \ell_{i} \bar\ell_{j}$.
In optimum case, it can be written as \cite{photon collider}
\begin{eqnarray}
\sigma_{\gamma \gamma \to \ell_i \bar\ell_j}(s)&=&
  \int_{\sqrt{a}}^{x_{max}}2zdz\hat{\sigma}_{\gamma \gamma \to \ell_{i} \bar\ell_{j}}
  (s_{\gamma\gamma}=z^2s) \int_{z^{2/x_{max}}}^{x_{max}}\frac{dx}{x}
 F_{\gamma/e}(x)F_{\gamma/e}(\frac{z^{2}}{x})
\end{eqnarray}
where $F_{\gamma/e}$ denotes the energy spectrum of the back-scattered photon for the
unpolarized initial electron and laser photon beams given by
\begin{eqnarray}
F_{\gamma/e}(x)&=&\frac{1}{D(\xi)}\left[1-x+\frac{1}{1-x}-\frac{4x}{\xi(1-x)}
  +\frac{4x^{2}}{\xi^{2}(1-x)^{2}}\right]
\end{eqnarray}
with
\begin{eqnarray}
D(\xi)&=&(1-\frac{4}{\xi}-\frac{8}{\xi^{2}})\ln(1+\xi)
  +\frac{1}{2}+\frac{8}{\xi}-\frac{1}{2(1+\xi)^{2}}.
\end{eqnarray}
Here $\xi=4E_{e}E_{0}/m_{e}^{2}$ ($E_{e}$ is the incident electron
energy and $E_{0}$ is the initial laser photon energy) and
$x=E/E_{E}$ with $E$ being the energy of the scattered photon moving
along the initial electron direction. The definitions of parameters
$\xi$, $D(\xi)$ and $x_{max}$ can be found in Ref.\cite{photon
collider}. In our numerical calculation, we choose $\xi=4.8$,
$D(\xi)=1.83$ and $x_{max}=0.83$.

\section{Numerical Results and Discussions}
As for the involved SM parameter, we take \cite{pdg}
\begin{eqnarray}
m_{\mu}=0.106{\rm ~GeV}, m_{\tau}=1.777{\rm ~GeV}, m_{b}=4.2{\rm ~GeV},
\alpha=1/137,\sin^2\theta_W=0.223
\end{eqnarray}

The TC2 parameters concerned in this process are $K_{\tau e}$,
$K_{\tau\mu}$, $K_{e\mu}$, $K_1$ and the mass of the extra gauge
boson $M_Z'$. $K_{e\mu}$ is very small, about $10^{-3}$, we will not
consider the $e-\mu$ conversion processes. In our calculation, we
have assumed $K_{\tau\mu}=K_{\tau e} \simeq \lambda \simeq 0.22$
 \cite{exp-tc2,tc2-cla}, which $\lambda$ is the Wolfenstein
parameter \cite{Wolfenstein}. It has been shown that the vacuum
tilting (the topcolor interactions only condense the top quark but
not the bottom quark), the coupling constant $K_{1}$ should satisfy
certain constraint, i.e. $K_{1}\leq 1$ \cite{K_1}. The limits on the
$Z'$ mass $M_Z'$ can be obtained via studying its effects on various
experimental observables \cite{exp-tc2}. Ref.\cite{m_z'}, for
example, has been shown that to fit the electroweak mearsurement
data, the $Z'$ mass $M_Z'$ must be larger than $1$
 TeV. As numerical estimation, we choose the center-of-mass
energy $\sqrt{s}=500$ and $1000$ GeV, to observe the different
behavior in the two energy area, and take the $M_Z'$ and $K_1$ as
free parameters. Finally, Note that the charge conjugate $\bar \tau
\mu(e)$ production channel are also included in our numerical study.

\def\figsubcap#1{\par\noindent\centering\footnotesize(#1)}
\begin{figure}[bht]%
\label{modes}
\begin{center}
\hspace{-0.25cm}
 \parbox{6.05cm}{\label{modes}\epsfig{figure=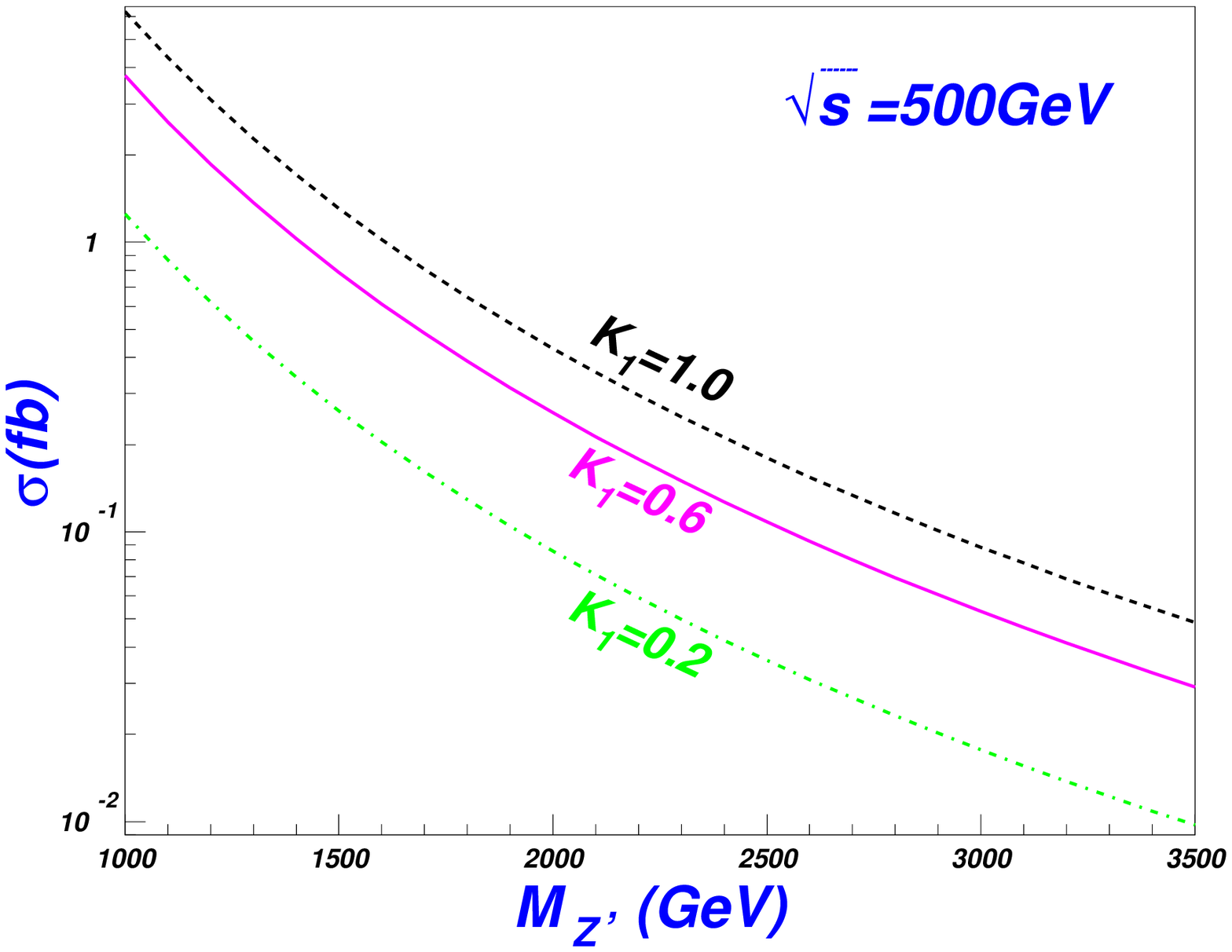,width=6.25cm}
 \figsubcap{a}\label{modes}}
 \hspace*{0.2cm}
 \parbox{6.05cm}{\epsfig{figure=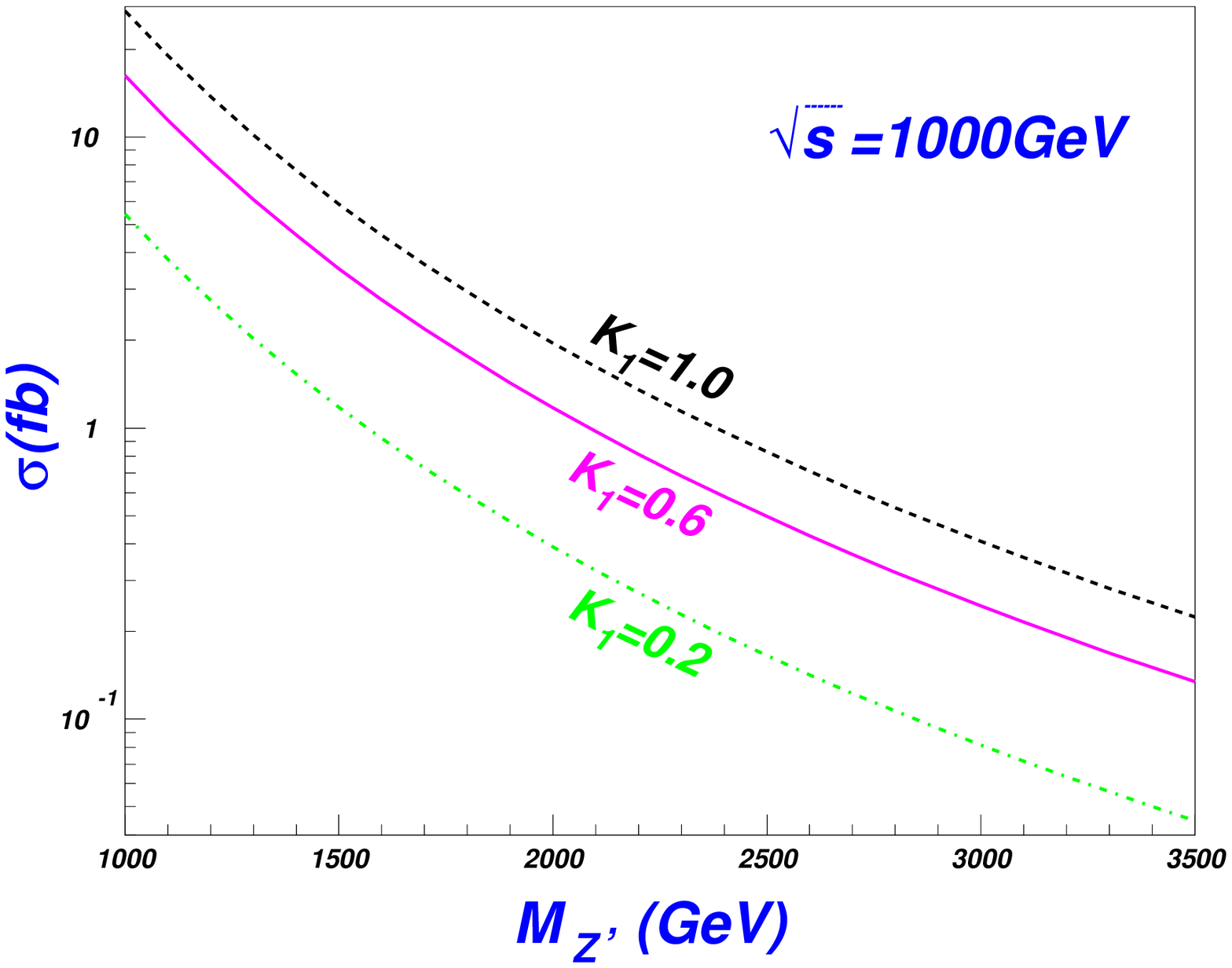,width=6.25cm}
 \figsubcap{b}\label{masses}}
 \caption{ The cross section $\sigma$ of the LFV process $\gamma\gamma
            \to\tau\bar\mu(\bar e)$ as a function of the gauge boson $Z'$
             mass $M_{Z'}$ for  $K_{1}=0.2$, $0.6$ and $1.0$ with
             (a) $\sqrt{s}=500GeV$ (b)$\sqrt{s}=1000GeV$. \label{fig2} }
\end{center}
\end{figure}

\def\figsubcap#1{\par\noindent\centering\footnotesize(#1)}
\begin{figure}[bht]%
\label{modes}
\begin{center}
\hspace{-0.25cm}
 \parbox{6.05cm}{\label{modes}\epsfig{figure=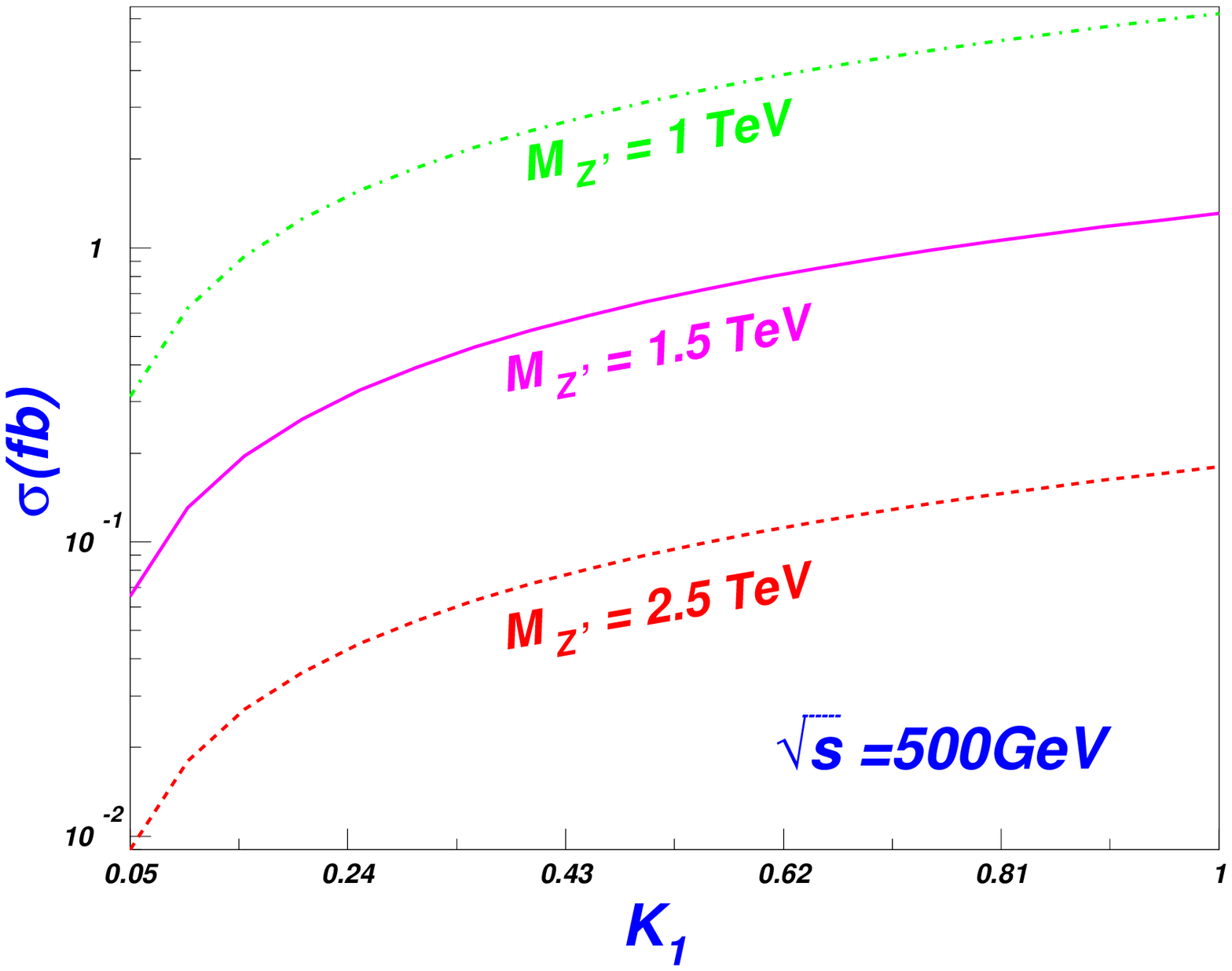,width=6.25cm}
 \figsubcap{a}\label{modes}}
 \hspace*{0.2cm}
 \parbox{6.05cm}{\epsfig{figure=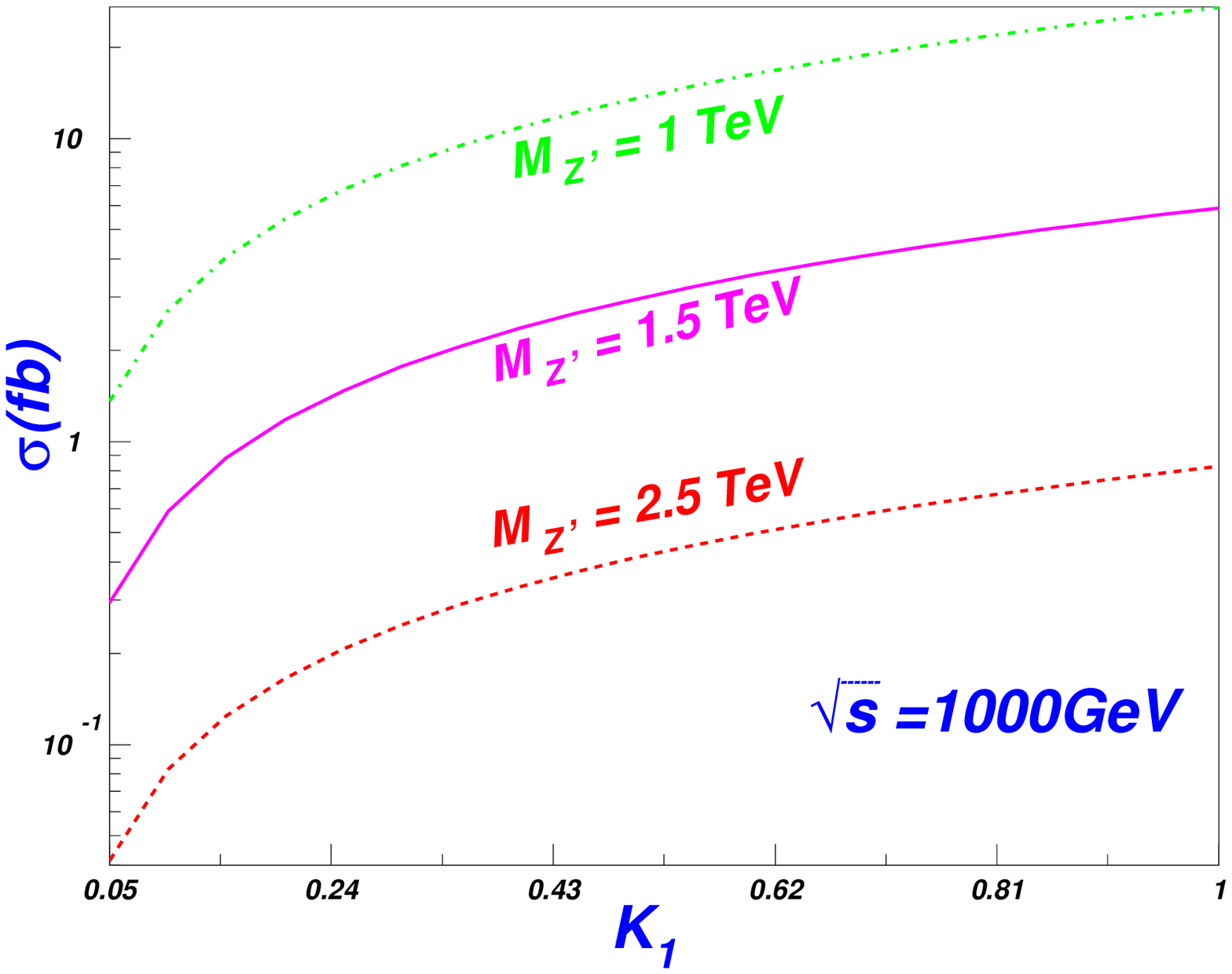,width=6.25cm}
 \figsubcap{b}\label{masses}}
 \caption{The cross section $\sigma$ of the LFV process $\gamma\gamma
            \to\tau\bar\mu(\bar e)$ as a function of the parameter $K_1$
with the gauge boson $Z'$ mass $M_{Z'}=1$ , $1.5$ and $2.5$ TeV for
(a) $\sqrt{s}=500GeV$ (b)$\sqrt{s}=1000GeV$. \label{fig3}  }
\label{modmass}
\end{center}
\end{figure}

In Fig.~\ref{fig2} we plot the production cross  section $\sigma$ of
the LFV process $\gamma\gamma \to \ell_i \bar\ell_j$ as a function
of $M_{Z}$ for three values of the parameter $K_{1}$: $K_{1}=0.2$,
$0.6$, and $1.0$. We can see from Fig.2 that the production cross
section $\sigma$ increases as $K_{1}$ increasing and strongly
suppressed by large $M_{Z'}$. This situation
 is slightly different from
the result of $e^+e^-\to \ell_i \bar\ell_j$ in \cite{eemutau-tc2},
in which from Fig.1 we can see the cross section of $e^+e^-\to
\bar\mu\tau$ increases with $K_1$ decreasing.  The reason is that
the $Z'\tau\bar\tau$ coupling involved in the process $\gamma\gamma
\to \tau\ell_i$($\ell_i=e$ or $\mu $) is proportional to
$1/\tan\theta^\prime\sim \sqrt{K_1}$, while the $Z'e^+e^-$ contains
$\tan\theta^\prime\sim \frac{1}{\sqrt{K_1}}$ and $\tan\theta'<<1$.
We can feel from this point the spirit of the technicolor models: to
give the natural top quark mass, the third generation is singled out
from the former two ones, so that it always shows distinct features.

The background for $\gamma \gamma \to e \bar{\tau}$ comes from
$\gamma\gamma \to \tau^{+}\tau^{-} \to
\tau^{-}\nu_{e}\bar{\nu}_{\tau}e^{+}$, $\gamma\gamma \to W^{+}W^{-}
\to \tau^{-}\nu_{e}\bar{\nu}_{\tau}e^{+}$ and $\gamma\gamma \to
e^{+}e^{-}\tau^{+}\tau^{-}$ \cite{rrmutau-susy}, and we make
kinematical cuts \cite{l-prod-ILC}: $|\cos\theta_\ell|<0.9$ and
$p^{\ell}_{T}>20{\rm ~GeV}$ ($\ell=e,\mu$), to enhance the ratio of
signal to background. With these cuts, the background cross sections
from $\gamma\gamma \to \tau^{+}\tau^{-} \to
\tau^{-}\nu_{e}\bar{\nu}_{\tau}e^{+}$, $\gamma\gamma \to W^{+}W^{-}
\to \tau^{-}\nu_{e}\bar{\nu}_{\tau}e^{+}$ and $\gamma\gamma \to
e^{+}e^{-}\tau^{+}\tau^{-}$ at $\sqrt{s}=500$ GeV are suppressed
respectively to $9.7\times 10^{-4}$ fb, $1.0\times 10^{-1}$ fb and
$2.4\times 10^{-2}$ fb (see Table I of \cite{l-prod-ILC}). To get
the $3 \sigma$ observing sensitivity with $3.45 \times 10^2$
fb$^{-1}$ integrated luminosity \cite{tesla}, the production rates
of $\gamma \gamma \to \tau\bar{e}, \tau \bar{\mu}$ after the cuts
must be larger than $2.5\times 10^{-2}$ fb \cite{l-prod-ILC}. We see
from Fig.\ref{fig2} that under the current bounds from $\ell_i \to
\ell_j \gamma$\cite{exp4-l-lr} and $\mu \to 3 e$\cite{z'couple}, the
LFV couplings in TC2 models can still large enough to enhance the
productions $\gamma\gamma \to e\bar{\tau}, \mu \bar{\tau}$ to the $3
\sigma$ sensitivity and may be detected in the future ILC colliders.
Finally note that we in Fig.2 only show the results of the channels
with the $\tau$ lepton in the final states, i.e., $\gamma\gamma \to
\tau\bar{\mu}$, $\tau\bar{e}$.

Fig.3 shows that the cross section of the LFV processes as a
function of $K_1$ for $ M_Z'= 1$, $1.5$,
 $2.5$ TeV. We can see more clearly that the cross
 section is increasing as the $K_1$ increasing.

\def\figsubcap#1{\par\noindent\centering\footnotesize(#1)}
\begin{figure}[bht]%
\label{modes}
\begin{center}
\hspace{-0.25cm}
 \parbox{6.05cm}{\epsfig{figure=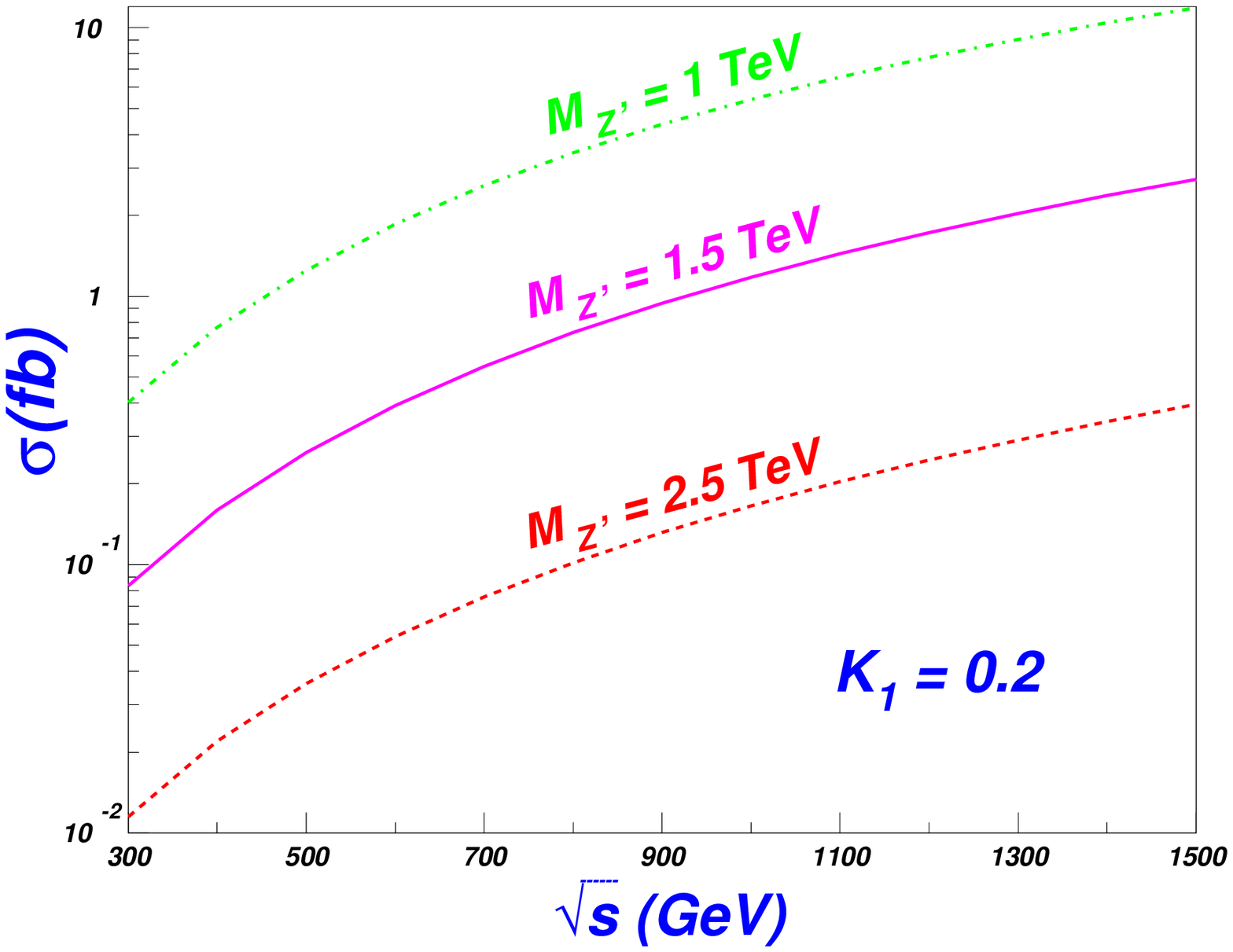,width=6.25cm}
 \figsubcap{a}}
 \hspace*{0.2cm}
 \parbox{6.05cm}{\epsfig{figure=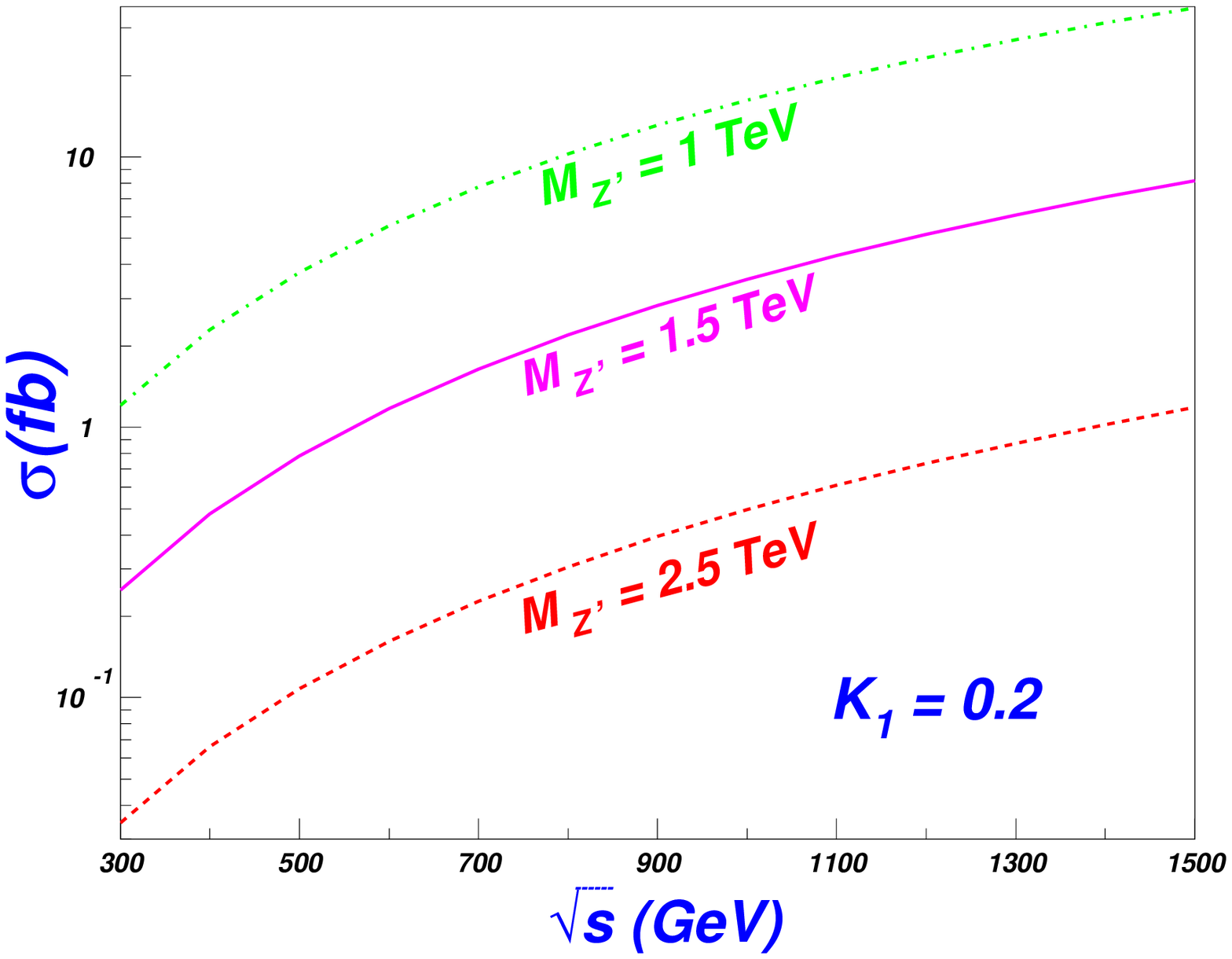,width=6.25cm}
 \figsubcap{b}}
 \caption{The dependence of the cross section $\sigma$ of the LFV process
$\gamma\gamma \to\tau\bar\mu(\bar e)$ on the center-of-mass energy
$\sqrt{s}$ for  $M_{Z'}=1$, $1.5$, and $2.5$ TeV with (a)
$k_{1}=0.2$, (b) $k_{1}=0.6$.  \label{fig4}  }
\end{center}
\end{figure}

We also show the cross sections of $\gamma \gamma \to \ell_i
\bar\ell_j$ as a function of center-of-mass energy $\sqrt{s}$ of the
ILC in Fig.4. We see that with the increasing of the center-of-mass
energy, the cross sections of these processes are not compressed,
instead of becoming larger. This is different with the results in
\cite{eemutau-tc2}, since, as mentioned above, the contribution of
the $\gamma\gamma \to \ell_i\bar\ell_j$ are the the results of t-
and u-channels, while in the processes $e^+e^- \to
\ell_i\bar\ell_j$,  the s-channel contribution
 decreases with the increasing $\sqrt{s}$ when the center-of-mass
 energy of the processes arrives at the critical value\cite{eemutau-tc2}.
Actually, we can also feel the larger cross section with larger
center of mass from Fig.2 and Fig.3.

\null \noindent {\small Table I: Theoretical predictions for the
$\ell_i\bar\ell_j$ ($i\neq j$) productions at $\gamma\gamma$
collision at the ILC. SUSY and TC2 predictions are the optimum
values. The collider energy is $500$ GeV.} \vspace*{0.1cm}
\begin{center}
\begin{tabular}{|l|l|l|}
\hline
 &~~SUSY ~~&~~~~TC2~~  \\
\hline ~~$\sigma(\gamma\gamma \to \tau\bar\mu)$  &~~${\cal O}
(10^{-2})$ fb & ~~~~$1$ fb \\ \hline ~~$\sigma(\gamma\gamma \to
\tau\bar e)$  &~~${\cal O} (<10^{-1})$ fb & ~~~~$1$ fb\\ \hline
~~$\sigma(\gamma\gamma \to \mu\bar e)$ ~~~~ &~~${\cal O} (<10^{-3})$
fb ~~~~&  ~~~~$10^{-3}$ fb~~~~\\ \hline
\end{tabular}
\end{center}

 As discussed in the former sections, motivated by the fact that
any process that is forbidden or strongly suppressed within the SM
constitutes a natural laboratory to search for any new physics
effects, LFV processes have been the subject of considerable
interest in the literature. It turns out that they may have large
cross sections, much larger than the SM ones, within some extended
theories such as the R-parity violating MSSM \cite{rrmutau-susy} and
the TC2 models. However, in the  R-parity violating MSSM, as
discussed in \cite{rrmutau-susy},
 the LFV coupling by the exchange of the squark
is $\lambda_{ijk}\sim 10^{-2}$, much smaller than that of the TC2
models ($ K_{\tau\mu(e)}\sim~ 0.2$). Therefore we can evaluate that
in the SUSY models the sigma of the LFV process $\gamma\gamma \to
\ell_i\ell_j$ is about $2-3$ order smaller than that in the TC2
models, as shown in table. I.

\section{Conclusion}
We have performed an analysis for the TC2-induced LFV productions of
$\tau\bar\mu$ and $\tau \bar e$ via $\gamma \gamma$ collision at the
ILC. We found that in the optimum part of the parameter space, the
production rate of $\gamma \gamma \to \tau\bar{\mu}(\bar e)$ can
reach $1$ fb. This means that we may have $100$ events each year for
the designed luminosity of $100$ fb$^{-1}$/year at the ILC. Since
the SM value of the production rate is completely negligible, the
observation of such $ \tau\bar{\mu}(\bar e)$ events would be a
robust evidence of TC2. Therefore, these LFV processes may serve as
a sensitive probe of TC2.

\vspace{2.5mm}
 {\bf \large Acknowledgement}
\vspace{2.5mm}

  We would like to thank J. J. Cao, J. M. Yang
and C. P. Yuan for helpful discussion.

\end{document}